\documentclass[12pt]{iopart}
\input{epsf.sty}
\usepackage{multicol}
\usepackage{epsfig,bbm,bm}
\usepackage{amssymb}

\def\beq{\begin{equation}}
\def\eeq{\end{equation}}
\def\beqa{\begin{eqnarray}}
\def\eeqa{\end{eqnarray}}
\def\ba{\begin{eqnarray}}
\def\ea{\end{eqnarray}}
\def\be{\begin{equation}}
\def\ee{\end{equation}}
\def\r{\rho}

\def\p{\phi}

\def\s{\psi}
\def\T{\tau}

\def\DD{D \bar D}

\def\pprime{\prime\prime}
\def\ap{\alpha^{\prime}}

\begin{document}

\title[Inflation]{Slow Roll in Brane Inflation}
\author{Sarah Shandera\dag}
\address{\dag\ Laboratory for Elementary Particle Physics\\ 
Cornell University\\ 
  Ithaca, NY 14853}
  \ead{seb56@mail.lns.cornell.edu}

\begin{abstract}
We address the issue of slow-roll in string theory models of inflation. Using a K\"{a}hler transformation and results from the D3-D7 model, we show why we expect flat directions to be present and slow-roll to be possible in general. We connect with earlier discussions of shift symmetry for $T^6/Z_2$ and $K3\times T^2/Z_2$ compactifications. We also collect various contributions to the inflationary potential and discuss their importance for slow-roll. We include a few simple checks of the form of the Kahler potential on $T^6/Z_2$ using T-duality.
\end{abstract}

\submitto{JCAP}
\maketitle
\date{\today}
\section{Introduction}
Models of stringy inflation have matured considerably in the past year as mechanisms of moduli stabilization have been found and implemented, and aspects of specific warped compactifications have been studied. We currently have a novel situation in string theory: these scenarios not only provide an interesting arena to study moduli and geometry, but are guided by close connections with observation. In particular, the models should satisfy the slow-roll conditions and agree with the observed spectrum of density fluctuations. They must include the standard model and a mechanism for reheating. They may also give testable predictions for details of the CMB spectrum that are just beyond current observational reach and perhaps for phenomena such as cosmic strings. Combining all of this in a single framework is a tall order, and there are many suggestions in the literature. We focus here on the first big hurdle: slow-roll. We will show how the physical picture is clarified by using a form of the potential that is invariant under K\"{a}hler transformations. We show that there is a very general argument for why we expect flat directions in the inflaton potential, and give details of two geometries: $T^6/Z_2$, where the geometric picture is simple, and the more realistic $K3\times T^2/Z_2$, where many details have already been worked out. We will also summarize the many contributions to the inflationary potential and give a few simple calculations using T-duality.

In the early days, work on brane inflation \cite{Dvali:1998pa} focused on achieving a flat enough potential between a non-BPS brane pair to allow for about 60 e-folds of inflation. This work was done assuming all moduli were stabilized in a way that did not affect the inflationary dynamics. At first glance, it seemed slow-roll could not be achieved unless a brane and an antibrane were separated by a distance greater than the size of the compactification. However, when the fact that the branes were sitting in a compact space was accounted for, slow-roll was found either by tuning the brane-antibrane case or by using branes at angles \cite{Burgess:2001fx}, \cite{Garcia-Bellido:2001ky}, \cite{Jones:2002cv}, \cite{Buchan:2003gx}. In a complete string theory picture, the inter-brane potential is easily made small by introducing warped throats in the compactification manifold, but other contributions to the inflationary potential complicate the question. In particular, recent proposals that solve the problem of moduli stabilization show contributions that appear to ruin the slow-roll. But as we will discuss, there are many other terms in the potential that must also be calculated and flat directions that can be found.

The model that has emerged as a successful general framework for brane inflation is developed in three works, which we refer to as GKP\cite{Giddings:2001yu} (Giddings, Kachru and Polchinski), KKLT \cite{Kachru:2003aw} (Kachru, Kallosh, Linde and Trivedi) and KKLMMT \cite{Kachru:2003sx} (Kachru, Kallosh, Linde, Maldacena, McAllister 
and Trivedi). The set-up is IIB string theory compactified on a Calabi-Yau 3-fold with one or more warped throats. Moduli are stabilized by fluxes and non-perturbative effects\footnote{Moduli fixing is a critical part of any fully consistent inflation model, but we will not review it here. For a general discussion, see for example \cite{Silverstein:2004id}.} and the warping of the throat aids inflation. Including multiple warped throats allows more freedom in setting multiple scales. Mobile D3 branes and D7 branes can be present in the bulk of the CY, and $\bar{D3}$ branes naturally sit at the IR tips of the throats. The $\bar{D3}$ branes lift the AdS vacuum to a meta-stable dS minimum. It is clear by now that the positions of $D7$-branes in a flux-stabilized N= 1 vacuum are fixed (though there are flat directions in N= 2 vacua) \cite{D'Auria:2004qv}.
This means the only candidate for the inflaton in brane inflation in the
KKLMMT like scenario is the position of a $D3$-brane. If the $D3$-brane 
annihilates with a $\bar {D3}$-brane towards the end of inflation, 
this is brane-antibrane inflation.
If the $D3$-brane collides with the $D7$-branes towards the end of 
inflation, this is the $D3-D7$ inflationary scenario. Since the 
attractive force in either scenario is, by itself, weak enough for 
slow roll, it is the other contributions to the inflaton 
mass that dictate the slow-roll condition. In this sense, the slow-roll 
analysis for these two scenarios is essentially identical. If the 
$D3$-brane position has an approximate shift symmetry (flat direction) at some 
particular locations in the CY manifold, then the relative forces
($D3-D7$ versus $D3-{\bar {D3}}$) will determine which scenario will 
take place. Most effects we will discuss apply equally to both calculations.

The KKLMMT model at first glance appears to suffer from the $\eta$-problem. That is, it does not find a small enough slow-roll parameter 
\be
\eta = M_{Pl}^2\frac{V^{\pprime}}{V}
\ee
where $V^{\pprime}$ is the second derivative of the potential with respect to the inflaton. The other slow-roll parameter, $\epsilon$, is defined by
\be
\epsilon=\frac{M_{Pl}}{2}\left(\frac{V^{\prime}}{V}\right)^2
\ee
where we want $\epsilon\ll1$. Since $\epsilon$ is typically the smaller of the two parameters, we shall focus on $\eta$ to determine the slow roll condition. To calculate $\eta$ the derivatives must be taken with respect to the canonical inflaton. To have enough inflation to solve the horizon, the defect, the flatness and the angular momentum problems, typically $|\eta| < 1/50$. But in KKLMMT, the contribution from the K\"{a}hler potential alone gives $|\eta|\sim2/3$. 

However, there are in general many contributions to the inflationary potential.
To determine slow-roll, we need to know all the pieces that contribute more than 1\% (since $V^{\prime\prime}/V\sim1/60$). As a first pass at listing the contributions to the potential, we can write
\be
\label{listterms}
V = V^{F}+V^{D}+V_{SB}
\ee
where the last term includes effects which break supersymmetry. In general, there may be many contributions to each piece. The F-term, for example, contains the superpotential from fluxes ($W_{0}$) and non-perturbative effects ($W_{np}$) and the K\"ahler potential ($K$). Supersymmetry breaking terms will include interactions between branes, antibranes and branes with fluxes. Writing out the F-term and writing some explicit contributions to $V_{SB}$, we have
\be
V = e^{K}(g^{a\bar b}D_aW\overline{D_bW} - 3|W|^2) + V^{D} + V_{(\bar D)} + V_{(\DD)} + \dots
\ee 
where
\be
W = W_{0}+W_{np}
\ee
$V_{(\bar D)}$ comes from $\bar{D3}$ branes in the throat, and $V_{(\DD)}$ is the interbrane potential. The inclusion of the D-term may be motivated by considering flux on D7 branes \cite{Kallosh:2003ux} or by trying to include the standard model \cite{Burgess:2004kv}. Note that these are no-scale models; that is, when the F-term is evaluated, we have 
\be
V^F = e^{K}(g^{a\bar b}D_aW\bar{D_bW} - 3|W|^2)\rightarrow e^K(\Sigma_{i,j\neq\rho}g^{i\bar{j}}D_iW\bar{D_jW})
\ee
This cancellation means we will have to find some different mechanism (i.e., non-perturbative effects) to fix the K\"{a}hler moduli.

In the F-term, $W_{0}$ can be fixed to a small constant in the KKLMMT model. The non-perturbative superpotential (e.g. from gaugino condensation \cite{Kachru:2003aw}) has the form
\be
\label{Wnp}
W_{np}=Ae^{-a\sigma}
\ee 
where $\sigma$ is the imaginary part of the K\"{a}hler modulus, $\rho$. The original KKLT paper made general arguments for why we expect this non-perturbative superpotential, but it is checked in detail in \cite{Gorlich:2004qm}.

The term $V_{(\bar D)}$ lifts the minimum from AdS to dS and comes from adding $\bar{D3}$ branes in the tip of the throat. It has the form
\be
V_{(\bar D)}=\frac{D}{\sigma^2}
\ee
where $D$ is a postitive constant. 

Taking as an example the case of D3-$\bar{D3}$ interaction, the term $V_{(\DD)}$ has the form (ignoring warping for now)
\be
\label{DDbar}
V_{(\DD)}=2T_3\left(1-\frac{A}{y^4}\right)
\ee
where $y$ is the separation of the branes and $A$ is a constant that depends on brane tension and the compactification volume. In the warped case, the distance to the tip of the throat is measured by the coordinate $r$. The $\bar{D3}$ is located at $r_0$, close to the tip, and the $D3$ is at $r_1$. The warped coordinate is
\be
R^4 = r^4h(r)
\ee
where
\be
h(r) = e^{-4A}
\ee
The warped throat flattens the inter-brane potential, modifying Eq.(\ref{DDbar}):
\be
V_{(D\bar{D})} = 2T_3\frac{r_0^4}{R^4}\left(1-\frac{1}{N}\frac{r_0^4}{r_1^4}\right)
\ee
where $N$ is the number of D3 branes.

Some Mathematica code has been developed (and is freely available) based on the D3/D7 model. It is described in \cite{Kallosh:2004rs}. The code is useful for following the dynamics of the scalar fields and computing cosmological consequences, but does not contain all the physics of Eq.(\ref{listterms}).

It may be possible to avoid the lengthy calculations suggested by Eq.(\ref{listterms}) if there is a symmetry that provides a nearly flat direction for the inflaton. It was noted in Appendix F of KKLMMT that if the superpotential ($W$) also had a dependence on the inflaton, it might cancel the K\"{a}hler contribution and allow slow roll. Several arguments were made for why just such a dependence might be reasonable. In \cite{Firouzjahi:2003zy}, it was shown that the superpotential preserves part of a shift symmetry of $\phi$ that is apparent by looking at the form of the physical volume. For the D3/D7 model, a similar argument was made in \cite{Hsu:2003cy} for the case of a mobile D7, which is found in $N=2$ theories but not in $N=1$. Using the special geometry construction of $K3\times T^2/Z_2$ \cite{D'Auria:2004qv} with moduli for the postitions of D3 branes (see \S2 for details), \cite{Hsu:2004hi} showed that the superpotential and K\"{a}hler potential still allow for a shift of the D3 brane position for the N=1 case. Furthermore, the authors of \cite{Berg:2004ek} (and a shorter version \cite{Berg:2004sj}) argue that the non-perturbative superpotential must include dependence on $\phi$ after one-loop contributions are included. In fact, we can immediately see with a K\"{a}hler transformation why $W$ can depend on $\phi$. This argument holds to a first approximation in a model-independent way. The only case that has been worked out in detail, $K3\times T^2/Z_2$, clearly has flat directions, so these two statements together are good evidence that such symmetries may be generic. The ideas are similar in spirit to earlier work done in \cite{Freese:1990mu}, \cite{Adams:1992bn}, and \cite{Freese:1990rb}, which considered the inflaton as a pseudo-Goldstone boson.

Let us see how this works. Consider the F-term potential
\be
\label{fterm}
V^{F} = e^{K}(g^{a\bar b}D_aW\overline{D_bW} - 3|W|^2)
\ee 
where
\be
D_aW = \partial_aW+W\partial_aK
\ee
and $a$ and $b$ label moduli. We will make use of the combination
\be
\label{invar}
G = K + \log|W|^2.
\ee
This is invariant under K\"{a}hler transformations
\be
K\rightarrow K+f(\phi)+\overline{f(\phi)}\hspace{0.3in}W\rightarrow We^{-f(\phi)}
\ee
Then Eq.(\ref{fterm}) can be rewritten in terms of $G$ as
\be
V^F=e^G\left[G^{a\bar{b}}\frac{\partial G}{\partial\phi_a}\overline{\frac{\partial G}{\partial\phi_b}}-3\right]
\ee
Since the F-term potential depends only on $G$, dependence on $\phi$ can be shifted between the superpotential and K\"{a}hler potential. 

Let us begin with the simplest form of the K\"{a}hler potential that is frequently assumed \cite{DeWolfe:2002nn}. Our guide for how the inflaton $\phi$ should appear in $K$ is the kinetic term in the action:
\be
$\cal{L}$_{kinetic}=-\frac{\partial^2 K(\phi,\bar{\phi})}{\partial\phi^i\partial\bar{\phi}^j}\partial_{\mu}\phi_i\partial^{\mu}\bar{\phi}^j
\ee
Since this depends on the mixed derivative, it does not fix terms that depend only on $\phi$ or $\bar{\phi}$. From this, we find
\be
\label{K1}
K_1(\phi,\bar{\phi},\rho,\bar{\rho}) = -3\ln[\rho+\bar{\rho}-\phi\bar{\phi}]\approx -3\ln[\rho+\bar{\rho}]+3\frac{\phi\bar{\phi}}{\rho+\bar{\rho}}
\ee
This form appears to treat the real and imaginary parts of $\phi$ equally. Let us compare with the result in the special geometry construction on $K3\times T^2/Z_2$ \cite{D'Auria:2004qv}, where the form of $K$ depends only on the imaginary part of $\phi$. This is a geometric result, not an approximation. In our notation, it is
\be
\label{Imonly}
K_2(\phi,\bar{\phi},\rho,\bar{\rho}) = -3\ln\left[\rho+\bar{\rho}+\frac{(\phi-\bar{\phi})^2}{2}\right] 
\ee
Both $K_1$ and $K_2$ are equally valid from the perspective of kinetic terms. But since the real part of $\phi$ is not present in $K_2$, it will not receive a large mass in the manner of KKLMMT, and it appears we have a flat direction at this level. In a general brane inflationary scenario, there are several 
possible directions $\phi_i$ so that $\eta$ is really a matrix
\be
\eta_{ij} = M_{Pl}^2\frac{\partial_{i}\partial_{j}V}{V}
\ee
It is sufficient for slow roll if every $\eta_{ij} \ll 1$. However, 
we actually require a weaker constraint.
If one or more $\phi_i$ start to oscillate, their oscillations 
will be rapidly damped by inflation. So, in the diagonal basis, large 
positive eigenvalues of $\eta$ are totally harmless, provided they are 
stabilized, that is, the corresponding slow-roll parameter $\epsilon$ vanishes. For 
slow-roll inflation,
we require at least one eigenvalue of $\eta$ to be small. So we expect slow-roll if $Re(\phi)$ is protected. 

How can we explain the difference between Eq.(\ref{K1}) and Eq.(\ref{Imonly})? We will first determine a toy form of the superpotential for each case (ignoring volume stabilization) to motivate the form of the K\"{a}hler transfromation that connects $K_1$ and $K_2$. Throughout these calculations, we will assume $\rho$ is large (since we want the supergravity description to be valid) and that $Re(\rho)\gg Im(\rho)$ (since we want non-imaginary kinetic terms). We also take $\phi\ll\rho$ since the separation between branes should be smaller than the compactification size. As in \cite{Firouzjahi:2003zy}, we will enforce the supersymmetry conditions to determine a basic expression for the superpotential. This form may be modified when the K\"{a}hler modulus is stabilized, but it is all we need to see the relationship between the two possible K\"{a}hler potentials. We require
\ba
D_{\phi}W_{(s)}&=&\partial_{\phi}W_{(s)} + W_{(s)}\partial_{\phi}K=0\\\nonumber
D_{\rho}W_{(s)}&=&0
\ea
Using the K\"{a}hler potential in Eq.(\ref{K1}), with the inflaton outside the logarithm, and taking $\phi=\bar{\phi}$ and $\rho=\bar{\rho}$ (since $W$ should be holomorhpic) we find
\be
\label{W1}
W_{1(s)}=w_0\rho^{3/2}e^{\frac{-3\phi^2}{4\rho}}\sim w_0\rho^{3/2}\left(1-\frac{3\phi^2}{4\rho}\right)
\ee
To first order, this agrees with the expression found in \cite{Firouzjahi:2003zy}, but the two differ starting with the $\phi^4$ term. If we do the same using $K_2$, we find that to all orders
\be
\label{W2}
W_{2(s)}=w_0\rho^{3/2}.
\ee

To first order, we can relate the two K\"{a}hler potentials using a K\"{a}hler transformation. Comparing Eq.(\ref{W1}) and Eq.(\ref{W2}) suggests we choose:
\be
f(\phi)=-\frac{3\phi^2}{4\rho}
\ee
Applying this gives
\ba
\label{Ktrans}
W_{1(s)}&\rightarrow&w_0\rho^{3/2}\\\nonumber
K_1(\phi,\bar{\phi},\rho,\bar{\rho})&\rightarrow&-3\ln[\rho+\bar{\rho}]+3\frac{\phi\bar{\phi}}{\rho+\bar{\rho}}-\frac{3\phi^2}{4\rho}-\frac{3\bar{\phi}^2}{4\bar{\rho}}+\dots
\ea
Now, since we expect $Im(\rho)$ to be small compared to $Re(\rho)$ we can write
\ba
K_1(\phi,\bar{\phi},\rho,\bar{\rho})&\rightarrow&-3\ln[2Re(\rho)]-\frac{3(\phi-\bar{\phi})^2}{4Re{(\rho)}}+\dots\\\nonumber
&\approx&K_2
\ea

So we see that the pairs ($K_1,W_1$) and ($K_2,W_2$) are equivalent to first order. That is, the flat directions that are apparent from Eq.(\ref{Imonly}) are the same as the shift symmetries found in \cite{Firouzjahi:2003zy} and \cite{Hsu:2004hi}. Using Eq.(\ref{Imonly}) has the advantage of making the shift symmetry transparent, and is well motivated by the calculation carried out in the $K3\times T^2/Z_2$ case. However, at higher order we cannot connect $K_1$ and $K_2$ by a K\"{a}hler transformation because terms that are not holomorphic or antiholomorphic in $\phi$ appear. Also, the expansions of Eq.(\ref{K1}) and Eq.(\ref{Imonly}) have different coefficients for the $\phi^4/\rho^2$ term. Since $K_2$ is derived from geometry (and is consistent with finding $K$ from the Lagrangian \cite{D'Auria:2004qv}) and since $K_1$ (which is only a guess) can be rewritten in the same form, this indicates that the shift symmetry is in fact generically present. The differences at higher order should be resolved if we could determine the exact K\"{a}hler potential for $T^6/Z_2$ including brane position moduli. It is noted in \cite{D'Auria:2004qv} that this is geometrically related $K3\times T^2/Z_2$ without D7 branes. $N=$ 4 SUGRA for IIB on $T^6/Z_2$ is discussed in detail in \cite{D'Auria:2003jk}. Perhaps this leads to a geometrical determination of the K\"{a}hler potential that can be used similarly to Eq.(\ref{Imonly}) and would clear up the issues with higher order terms.

The rest of the paper is organized as follows. In \S2 we discuss a few details of the two known models: $K3\times T^2/Z_2$ and $T^6/Z_2$. In particular, we generalize previous discussions of shift symmetry to include more than one K\"{a}hler modulus and make use of T-duality to check the form of the K\"{a}hler potential.  In \S3 we comment on the relation between the analysis above and the many additional terms in the inflationary potential. We conclude in \S4.

\section{Specific Models}
\subsection{$K3\times T^2/Z_2$}
The $K3\times T^2/Z_2$ model is the only realistic case that has been studied in much detail. A great deal of work has been done to investigate the supersymmetric vacua in the presence of 3-form fluxes \cite{Andrianopoli:2003jf}, \cite{Tripathy:2002qw} and to include D3 and D7 branes to connect to the D3-D7 inflationary scenario \cite{D'Auria:2004qv}, \cite{Angelantonj:2003zx}. Since these results contain all the pieces needed for a scenario that allows D3/D7 or D3/$\bar{D3}$ inflation, we will review them here. For continuity with this work, we use the conventions of \cite{Andrianopoli:2003jf} in this section. 

There are three moduli to consider in addition to the positions of the D3 and D7 branes: the K\"{a}hler modulus $s$, the $T^2$ complex structure $t$, and the axion-dilaton $u$. These are defined as
\ba
s &=& C_{(4)}-iVol(K3)\\\nonumber
t &=& \frac{g_{12}}{g_{22}} + i\frac{\sqrt{det(g)}}{g_{22}}\\\nonumber
u &=& C_{(0)} + ie^{\Phi} 
\ea
where $g_{ij}$ is the metric on $T^2$. Notice that $s= -i\rho$ to compare with the KKLMMT notation. The position of the $r$th D3 brane along the torus $T^2$ is $\phi^r$, and the position of the $k$th D7 brane is $x^k$. The K\"{a}hler potential is \cite{Andrianopoli:2003jf}
\ba
\label{K3Kahler}
K &=& -\log\left[-8\left(Im(s)Im(t)Im(u)-\frac{1}{2}Im(s)(Im(x)^i)^2\right.\right.\\\nonumber
&&\left.\left.-\frac{1}{2}Im(u)(Im(\phi)^r)^2\right)\right]
\ea
This depends only on the imaginary part of the D3 brane position as discussed above. The scalar potential is
\ba
\label{scalarpot}
V &=& 4e^{2\Phi}e^{K}\left[\sum_{\Lambda=0}^{3}(g_{\Lambda})^2|X^{\Lambda}|^2\right.\\\nonumber
&&+\frac{1}{2}(g_0^2+g_1^2)(t-\bar{t})\left((u-\bar{u})-\frac{1}{2}\frac{(x^k-\bar{x}^k)^2}{(t-\bar{t})}\right)\\\nonumber
&&\left.+\frac{(\phi^r-\bar{\phi}^r)}{8(s-\bar{s})(u-\bar{u})}(g_0^2(\bar{u}x^k-\bar{x}^ku)^2+g_1^2(x^k-\bar{x}^k)^2)\right]
\ea
where the $X^{\Lambda}$ are the components of the symplectic section
\ba
\Omega &=& (X^{\Lambda}, F_{\Lambda}=\partial F/\partial X^{\Lambda})\\\nonumber
X^0 &=& \frac{1}{\sqrt{2}}\left(1-tu+\frac{(x^k)^2}{2}\right)\\\nonumber
X^1&=&-\frac{t+u}{\sqrt{2}}\\\nonumber
X^2&=&-\frac{1}{\sqrt{2}}\left(1+tu-\frac{(x^k)^2}{2}\right)\\\nonumber
X^3&=&\frac{t-u}{\sqrt{2}}\\\nonumber
X^k&=&x^k\\\nonumber
X^r&=&\phi^r
\ea
$F$ is related to the prepotential $\cal{F}$ by $F(X^{\Lambda})/(X^0)^2=\cal{F}$. The $g_i$ are couplings related to the fluxes that break supersymmetry, so choosing various combinations of $g_i$ non-zero corresponds to different amounts of supersymmetry. In the $\cal{N}=$ 2 case, $g_0=g_1=0$ and any dependence on $\phi^r$ comes only through the factor $e^K$. In that case, both the D7 and D3 positions are moduli. For $\cal{N}=$ 1, the postitions of the D7 branes are fixed. This can also be seen from the earlier results of GKP: since the D7's are wrapping some 4-cycles, then if the complex structure moduli are fixed the position of the D7s are fixed.

Specializing to the $N=$ 1 case ($x^k=0$) simplifies the expressions above. In particular, the scalar potential becomes:
\be
\label{scalarpot2}
V = 4e^{2\Phi}e^{K}\left[\sum_{\Lambda=0}^{3}(g_{\Lambda})^2|X^{\Lambda}|^2+\frac{1}{2}(g_0^2+g_1^2)(t-\bar{t})(u-\bar{u})\right]
\ee
Note that Eq.(\ref{scalarpot2}) does not depend on the position of the D3 brane except through the K\"{a}hler potential. When the full potential depends on the inflaton only through the exponential factor $e^K$, with $K$ given as in Eq.(\ref{K3Kahler}), the contribution of the F-term to $\eta$ is always small (assuming $y\ll s$). For the special case of $u=t=-i$, Eq.(\ref{scalarpot2}) is identically zero. We can then use Eq.(\ref{scalarpot2}) and Eq.(\ref{K3Kahler}) to calculate the superpotential for this case. We find that the superpotential is constant, and so does not depend on the brane position. So for the case of $N=1$, regardless of the values of the other moduli, the contribution of the F-term to $\eta$ does not destroy the slow-roll. Some additional analysis of the D3-D7 model can be found in \cite{Koyama:2003yc}. A recent catalogue of results for IIB orientifolds with 3-form fluxes and D3 and D7 branes is \cite{Lust:2004fi}.

\subsection{$T^6/Z_2$}
A simpler model to analyze is a compactification on $T^6/Z_2$. We have explained in the introduction how shift symmetry can be seen in this case. Here, we will also see that in this model T-duality is a good guide to the form of the K\"{a}hler potential, particularly in the $N=$2 case where mobile D7 branes may be present. We will work in a general scenario with three K\"{a}hler moduli, three complex strucutre moduli, and the dilaton. Previous work on this model includes Ref. \cite{Frey:2002hf}, which gives examples of solutions with D=4, $N=3$ supersymmetry and discusses dualities of these models. Ref. \cite{Kachru:2002he} finds choices of fluxes that give $N=3$ and $N=1$ minima and shows that the fluxes typically stabilize the dilaton, all complex structure moduli, and some K\"{a}hler moduli. This work also discusses when solutions with different values for the dilaton or complex structure moduli are actually physically distinct. A detailed supergravity analysis is done in \cite{D'Auria:2003jk}. Here, we would like to use this simple geometry to derive the form of the K\"{a}hler potential when all moduli are kept, check that the results are consistent with T-duality, and illustrate the shift symmetry.

In the following discussion we consider three square tori $T_1, T_2$ 
and $T_3$ with radii $R_1$, $R_2$ and $R_3$ respectively.
Our ansatz for the metric in the Einstein frame is
\ba
\label{metric}
ds^2=(R_1 R_2 R_3)^{-2}g_{\mu\nu}dx^{\mu}dx^{\nu}+\sum_{i}^{3}R_i^2\, 
\delta_{mn}\, dy^{m_i}dy^{n_i}~~~~~~~m,n=1,2
\ea
where $g_{\mu\nu}$ stands for the four dimensional metric in the Einstein frame.
Here we assume $x^{\mu}$ and $y^{m_i}$ are dimensionful such that 
$R_i$ are dimensionless and 
$y^{m_1}\equiv(y^1,y^2)$, $y^{m_2}\equiv(y^3,y^4)$ and $
 y^{m_3}\equiv(y^5,y^6)$ represent the coordinates on $T^2_1$, $T^2_2$ and $T^2_3$.
The prefactor $(R_1 R_2 R_3)^{-2}$ is due to the transition from string frame to Einstein frame.

Up to second derivatives, the K\"{a}hler potential is found from the kinetic energy of mobile branes in the DBI action. In general there can be D7 branes wrapping different tori. We label these D7$_i$, where $i$ indicates which torus is {\it not} wrapped. Complex scalars $\psi_{i}$ and $\phi_i$ give the transverse positions of the D7$_i$ and the D3-branes along the directions of the $i$-th torus. $F_{\mu \nu}=\partial_{\mu}A_{\nu}-\partial_{\nu}A_{\mu}$
and $F_{(i)\mu m_j}=\partial_{\mu}{A_{(i)}\,}_{m_j}$ are the real-valued 
gauge field strengths living on the D7$_i$-brane, while $F'_{\mu \nu}$ 
is that on the D3-brane.

The DBI action for one D3-brane and three D7-branes, up to second derivatives, is given by
\ba 
\label{DBI}
S=&-&T_7\frac{(2\pi \alpha')^2}{4}\int d^8 \, \xi\sqrt{-g}
\left[\sum_{i}d_{ijk}(R_jR_k)^2
F_{(i)\mu\nu}F_{(i)}^{\mu\nu}\right.\\\nonumber
&+&\left.\sum_{i,j}d_{ijk}(R_iR_j)^{-2}
F_{(i)\mu m_j}F_{(i)}^{\mu m_j}
\right]\nonumber\\
&-&\frac{T_7}{2}\int d^8 \, \xi\sqrt{-g}\, e^{\Phi}\sum_{i}
\partial_{\mu}\psi_i \partial^{\mu}\bar{\psi_i}
-T_3\frac{(2\pi \alpha')^2}{4}\int d^4 \, \xi\sqrt{-g}
\, e^{\Phi}F'_{\mu \nu}F'^{\mu \nu} 
\nonumber\\
&-&\frac{T_3}{2}\int d^4 \, \xi\sqrt{-g}\sum_{i}d_{ijk}(R_j R_k)^{-2}
\partial_{\mu}\phi_i \partial^{\mu}\bar{\phi_i}
\nonumber\\
\ea
where the quantity $d_{ijk}$ is defined such that $d_{ijk}=1$ for 
$(i,j,k)$ a permutation of (1,2,3) and 0 otherwise. 

The background K\"{a}hler potential without mobile branes is given by
\ba
K_0=-\mbox{ln}(\tau+\bar{\tau})-\sum_{i}^{3}\mbox{ln}(\rho_i+\bar{\rho_i})
\ea
where the complex scalar fields (chiral superfields) are defined by
\ba
\T&=&e^{-\Phi}+iC_{(0)}\nonumber\\
\r_i&=&(R_jR_k)^2+i\,b_i \,
\ea
where $C_{(0)}$ is the RR scalar and the $b_i$ are related to the RR four-form field $C_{(4)}$. Note that $\tau=-it$ to compare to the notation used in the $K3\times T^2/Z_2$ section.
The easiest way to see the definition of $\T$ and $\r_i$ in this particular
way is to look at the gauge kinetic energy for gauge fields living on the D3 and D7-branes. The practical advantage of defining $\T$ and $\r_i$ in this way 
is that the scalar kinetic energies are diagonalized. Here we will stick with this form (rather than transfroming to something of the form of $K_2$) since the T-duality is easier to see.

Reading off the kinetic energy for mobile D3/D7-branes will lead us to the
following generalization of the K\"{a}hler potential
\footnote{K as given in Eq(\ref{Kahler}) agrees with the K\"{a}hler potential
obtained from the perturbed DBI action up to second order in $\p_i$ and $\s_i$. 
Eq.(\ref{Kahler}) is a natural generalization of the perturbed 
K\"{a}hler potential such that it has SO(3) symmetry with respect to $\p_i$
when $\r_1=\r_2=\r_3$ and also it respects the T-duality. In our 
discussions we assume the inflaton $\p_i$ or $\s_i$ is small, so we need $K$ only up to second order in those fields. But as discussed in the introduction, it would
be an interesting exercise to see if Eq.(\ref{Kahler}) actually is the 
K\"{a}hler potential to all order in $\p_i$ and $\s_i$.}
\ba
\label{Kahler}
K=-\ln\left[\T\r_1\r_2\r_3\left(1-\sum_{i}^{3}
\frac{\p_i\bar{\p_i}}{\r_i+\bar{\r_i}}
-\sum_{i}^{3}\frac{\s_i\bar{\s_i}}{\T+\bar{\T}}
-\sum_{i,j}^{3}d_{ijk}\frac{C_{ij}{\bar{C}}_{ij}}{\r_k+\bar{\r_k}}\right) 
 \right]\ .
\ea
We have defined the complex fields $C_{ij}$ from the real Wilson lines 
$A_{(i)m_j}$ such that
\ba
\label{C}
C_{ij}\equiv\sum_{m=1}^{2}i^{m-1}A_{(i)m_j}\ .
\ea
For example $C_{12}$ describes gauge fields living on D7${_1}$
and having the components along $T^2_2$:
\ba
C_{12}={A_1\,}_{y_3}+i{A_1\,}_{y_4}
\ea

As a self-consistent check, we investigate the invariance of $K$ under 
T-duality. Consider the T-duality along tori T$_2$ and T$_3$. We have
\ba
\label{T-duality}
\T\leftrightarrow \r_1~~~~~&,&~~~~~\r_2\leftrightarrow \r_3~~~~~,~~~~~
D3 \leftrightarrow D7_1~~~~~,~~~~~D7_2 \leftrightarrow D7_3\nonumber\\
\p_1\leftrightarrow \s_1~~~~~&,&~~~~~\p_2\leftrightarrow {C_1}_2
~~~~~,~~~~~\p_3\leftrightarrow {C_1}_3\nonumber\\
\s_2\leftrightarrow{C_3}_2~~~~~&,&~~~~~\s_3\leftrightarrow{C_2}_3
~~~~~,~~~~~{C_2}_{1}\leftrightarrow{C_3}_1
\ea
One can explicitly verify that $K$ given by
Eq.(\ref{Kahler}) is invariant under the above transformations.
One interesting aspect of this T-duality is the interchange of $\T$ and $\r_1$.

We now turn to the question of shift symmetry. To simplify the 
situation and emphasize our main result for shift symmetry, 
we truncate the gauge fields ${C_i}_j$ from our discussions. 
As before, we take the limit where
$\p_i,\s_i\ll\r_i$. In this limit,
and neglecting ${C_i}_j$,  $K$ is given up to second order in $\p_i$ and $\s_i$ by
\ba
\label{perturbK}
K=-\ln(\T+\bar{\T})-\sum_{i}^{3}\ln(\r_i+\bar{\r_i}) 
+\sum_{i}^{3}\frac{\p_i\bar{\p_i}}{\r_i+\bar{\r_i}}
+\sum_{i}^{3}\frac{\s_i\bar{\s_i}}{\T+\bar{\T}}
\ea 
Of course, the K\"{a}hler potential in Eq.(\ref{perturbK})
is not invariant under the whole set of transformations in Eq.(\ref{T-duality}) because we ignored ${C_i}_j$.

The KKLMMT moduli stabilization procedure is easily generalized to the case at hand. The K\"{a}hler moduli $\r_1, \r_2$ and $\r_3$ are respectively stabilized by gauge fields living on the three stacks D7$_1$, D7$_2$ and D7$_3$. These nonperturbative potentials originate from gaugino condensation of super-YM 
SU(N$_i$) on the D7$_i$. The form of nonperturbative superpotential on 
D7$_i$ is given by
\ba
\label{Wnpmulti}
{W_{np,i}}\sim e^{-a_i/{{g_{YM}}_i}^2} 
\ea
where $a_i$ is a particular property of the gauge group living on 
the stack of D7$_i$, like $N_i$.
On the other hand, from the four dimensional effective action point of view
\ba
\label{gYM}
\frac{1}{{{g_{YM}}_i}^2}\sim e^{-\Phi}\mbox{Re}(\r_i) \ .
\ea
This implies that
\ba
\label{TdualW}
{W_{np,i}}\sim e^{-a_i  \T\r_i} \ .
\ea
Each $W_{np,i}$ will stabilize $\r_i$ separately. 

\begin{figure}[htb]
\begin{center}
\includegraphics[width=0.7\textwidth,angle=0]{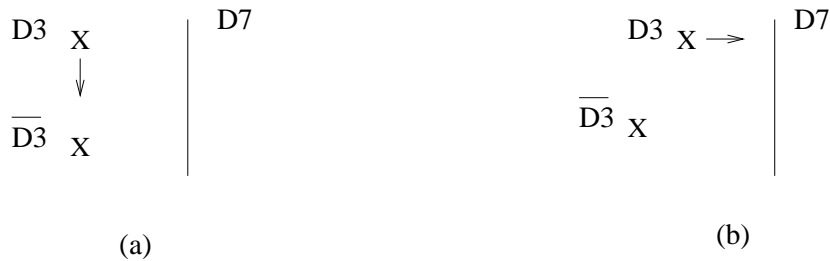} 
\caption{Simple possibilities for inflation. (a) The D3 moves along a flat direction, parallel to a D7 brane. (b) Whether $\DD$ or D3-D7 inflation occurs depends on the arrangement of branes.
\label{D3motion}}
\end{center}
\end{figure}

As we saw in the introduction, we can make a K\"{a}hler transformation to write the K\"{a}hler potential in a way that depends only on the real part of each $\phi_i$, so that we now have three flat directions. For the simple geometry of $T^6/Z_2$, we can in fact illustrate the symmetry. Figure \ref{D3motion}(a) shows a simple case of $\DD$ inflation where the D3 moves parallel to the D7. Note that this is an illustration of a flat direction of the F-term potential, not the interbrane potential. The geometric analogy is much more difficult to visualize for $K3\times T^2/Z_2$.

This fits nicely with the physical picture described in \cite{Firouzjahi:2003zy}, where the shift symmetry is understood by looking at the strong interaction between the mobile D3 and a heavy stack of D7 branes. The D3-brane can move around this stack of D7s either radially or azimuthally as it moves toward the $\bar{D3}$ in the throat. If we consider open strings stretched between the D3 and the stack of D7s, their mass must be proportional to the radial distance between the branes. As long as the D3 moves along the azimuthal direction with a fixed radial distance from the D7s, the strong interaction between them will remain unchanged and that is what we interpret as the flat direction of $V_F$.

We can also see from this picture that the locations of the D3, $\bar{D3}$ and D7 branes will decide which interaction drives inflation. For example, it is likely the D3 and $\bar{D3}$ will not be aligned exactly along a flat direction, as in Figure \ref{D3motion}(b). Then the D3-D7 interaction may be the source of the slow-roll. In a realistic case, we may expect to have several throats and several stacks of D7's, so there may be many competing effects. Just the competition between two throats can lead to an enhanced possibility for slow-roll \cite{Iizuka:2004ct}.

One final comment is in order regarding shift symmetry for inflation in the $N=2$ case. Consider a T-duality along tori T$_2$ and T$_3$. As mentioned before $\r_1\rightarrow \T$ and $\p_1\rightarrow \s_1$. The discussion of shift symmetry for $(\r_i,\p_1)$ is now mapped to $(\T,\s_1)$, and it is now the $\T$ stabilization which should be scrutinized. In fact, the situation seems more severe than before, because $\T$ is stabilized at the perturbative level. In
GKP it was shown that the perturbative superpotential has a linear 
dependence on $\T$, which will in general give $\eta \sim 1$.

To see this more explicitly, let us assume we have one  K\"{a}hler modulus, $\r$, and $\s_1$ is the inflaton. At the perturbative level $W=W(\r,\T)$\footnote{$\s$ appears only in cubic
form in non-perturbative superpotential \cite{Lust:2004fi} and we can safely ignore it
since $\s<<\r$.}.
With $K$ given as in (\ref{perturbK}) and with $\p=0$ (dropping the D3 brane), we obtain
\ba
\label{pertVF}
V_F&=&e^{K}(g^{a\bar b}D_a W \overline{D_b W}-3|W|^2)\nonumber\\
&\sim&2\T W_{\T}^2-2W W_{\T}+\frac{W^2}{2\T} 
+\frac{W^2}{2\T^2}\s_1\bar{\s_1} \ ,
\ea
where for simplicity we assumed $\T=\bar{\T}$ and $W=\bar W$.

Defining 
\ba
V_0=2\T W_{\T}^2-2W W_{\T}+\frac{W^2}{2\T} 
\ea
for $\T$ at the stable location, one sees
\ba
m_{\s_1}^2=\frac{W^2}{2\T} \ .
\ea
In the limit where $\T\gg1$ (since $\T=1/g_s$), $m_{s_1}^2\sim V_0$. For
the superpotential created by fluxes as in GKP with linear dependence
on $\T$, one can easily verify that $m_{s_1}^2\sim V_0$. This demonstrates that $\T$ stabilization would spoil the shift symmetry of $\s$ even at the perturbative level.

\section{Comments on the Inflationary Potential}
In this section we return to Eq.(\ref{listterms}) and the many contributions to the inflationary potential. The presence of a shift symmetry guarantees that the F-term will not ruin slow-roll, but we still must check the other terms. Also, in a model without shift symmetry, slow-roll may still take place if any large contributions are cancelled by other terms. At the end of this section we will review progress toward constructing complete models and possible observational consequences.

\subsection{$\ap$ Corrections}
Among the considerations for calculating the slow-roll potential are $\ap$ corrections. In general, it is important to ensure that any corrections to the superpotential be consistent with Eq.(\ref{invar}) and any known corrections to the K\"{a}hler potential. In other words, one should check that all results are invariant under K\"{a}hler transformations. 

Leading order $\ap$ corrections to the superpotential are calculated in \cite{Berg:2004ek} for the $T^2\times T^4/Z_2$ model and generalized to $T^6/Z_N$ (without moduli stabilization). A naive dimensional argument suggests a typical
dimensionless parameter involving $\ap$ is $\ap / L^2 $, where
$L$ is the size of the compactification manifold, which is typically 
a small number. So one might assume such corrections may be ignored. 
However, in specific KKLT like scenarios, there are additional 
scales generated in the vacuum, so one should check the 
size of the $\ap$ corrections.

In \cite{Berg:2004ek}, the correction to the gauge coupling was calculated by the background field method. In the presence of background gauge field strength $F$ with gauge field $A_i$
and Wilson lines ${\bf a}$, the one-loop contribution
(from the torus, the Klein bottle, the Mobious strip and
the cylinder) takes the form :
\ba
\Lambda_{1-loop} = \Lambda_0 + \frac{1}{8 \pi^2} F^2 \Lambda_2 ({\bf a})
\ea
where $\Lambda_0$ is a cosmological constant and $\Lambda_2 ({\bf a})$
is identified with 1-loop correction to the gauge coupling
\ba
g_{1-loop}^{-2} = g_{tree}^{-2} + \frac{1}{4\pi^2}\Lambda_2 ({\bf a})/\sqrt{-g}
\ea
In this way, the authors determined the 1-loop correction in the 9- and 5-brane
picture of Type II orientifold. After T-dualizing the 6 compact
dimensions, they obtain the 3- and 7-brane picture, where the
$A_i$ of a D9-brane becomes the position $\phi_i$ of a 3-brane,
which is identified as the inflaton. The result for the corrected gauge kinetic function on the D7 brane is
\be
{\it f}^7=-i\rho -\frac{1}{4\pi^2}\ln{\vartheta_1(\phi,\tau)}+ \dots
\ee
This gives a correction to the superpotential that can be expressed in terms of a function $w(\phi,\tau)$ as
\be
W_{np}\sim Ce^{-a{\it f}_7}=w(\phi,\tau)e^{ia\rho}
\ee
From here, the 1-loop contribution to the inflaton
mass (or equivalently the 1-loop contribution to $\eta$) can be determined. The corrected inflaton mass is
\beq
m^2 \simeq 2 H^2(1- \frac{V_{AdS}}{V_{dS}} \Delta)
\eeq
where $\Delta \simeq 0.1$. This matches the general form calculated in Appendix F of KKLMMT for the effect of introducing $\phi$ dependence in the superpotential. In KKLMMT the two scales are related by
\beq
V_{dS} = 3H^2 \simeq V_{AdS} + V_{\bar{D}} = V_{AdS}+\frac{D}{\sigma^2}
\eeq
Generically, we expect that $|V_{AdS}| > V_{dS}$ (since we want a small cosmological constant), so the $\ap$ correction can be quite significant. If $|V_{AdS}| \gg V_{dS}$, then higher order $\ap$ corrections to 
the superpotential as well as to the K\"{a}hler potential will be very 
important, and it is very questionable if they can ever be reliably 
determined. If $|V_{AdS}| \sim V_{dS}$, then next order $\ap$ 
correction will be important as well, since we want to determine
the slow-roll parameter to $1 \%$.

As we have seen, the inflaton potential (and so $\eta$) depends on the
particular combination $G$ of the K\"{a}hler potential $K$ and the
superpotential $W$. The calculation in \cite{Berg:2004ek} so far includes only the
1-loop contribution to $W$. Since a K\"{a}hler transformation can move
any contribution from $K$ to $W$ or vice versa, it will be important
to find the 1-loop contribution to $K$ and check to see if the contribution to $G$ or $\eta$ is indeed large. Also, the
calculation was carried out in simplified versions of the realistic
model. It is not entirely clear how valid extrapolation is to a
KKLMMT model.

K\"{a}hler corrections can usually be ignored for large volumes (and KKLT easily obtains a large volume, which is also needed for the supergravity description to hold), but \cite{Balasubramanian:2004uy} argues that the no-scale structure of the KKLT models makes the corrections significant for the qualitative form of the potential. Their concern is with a possible solution to the cosmological constant problem, and does not directly impact the $\eta$ question, but in general such corrections may be important. The form of perturbative corrections to the K\"{a}hler potential is worked out in \cite{Becker:2002nn}. The leading order is an $\alpha^{\prime3}$ term that depends on the dilaton and appears in the part of the K\"{a}hler potential that contains the K\"{a}hler modulus. Since the D3-brane moduli appear together with $\rho$, this changes the kinetic term for $\phi$ and so changes the rescaling that must be done to obtain the canonically normalized field. The form of non-perturbative corrections to the K\"{a}hler potential are discussed in \cite{Barreiro:1997rp}, \cite{Higaki:2003jt} and \cite{Binetruy:1996gw}.

\subsection{D-terms and brane interactions}
Investigations of how D-terms may be useful in inflation predate the brane inflation scenario \cite{Binetruy:1996xj}, \cite{Halyo:1996pp}, and continue to be discussed outside that framework \cite{Higaki:2004wc}, \cite{Kobayashi:2003rx}. We will review this work as it nicely illustrates the difference between D- and F-terms and since investigations along these lines may uncover ideas useful in brane inflation. 

The original idea of including D-terms was to solve the $\eta$ problem which arises from F-terms only. The issue is that all scalars receive a mass of order $m^2\sim H^2$, which ruins slow-roll. But if the D-term dominates and the inflaton is not charged under the gauge group involved, $\phi$ gets a mass only through loop effects. This can be small enough to keep $\eta\ll1$. This scenario goes by the name ``D-term inflation''. Despite the promise of small $\eta$, these models suffer from a runaway dilaton and are in numerical conflict with CMB measurements (the magnitude of the FI term is too large) \cite{Lyth:1998xn}. They also predict cosmic strings that would contribute too much to density perturbations. 

Much work has also been done to discuss D-terms in the brane inflation scenario. We will first outline some considerations that do not directly deal with the $\eta$ question, but arise any time D-terms are present. In \cite{Burgess:2003ic}, it was found that adding D-terms to the potential gave rise to dS vacua. Although several mechanisms have been used to achieve de Sitter space, starting with KKLMMT who included $\bar{D3}$ branes in the throat (and \cite{Saltman:2004sn} which demonstrated that $\bar{D3}s$ are not necessary), the advantage of adding D-terms is that the action is fully supersymmetric. The D3-D7 brane inflation model \cite{Dasgupta:2002ew} takes advantage of the benefits of D-term inflation to get slow-roll. This scenario has been well-established in the KKLMMT context by \cite{Hsu:2004hi}, \cite{Hsu:2003cy}, \cite{Dasgupta:2004dw}.

Furthermore, \cite{Burgess:2004kv} found that when the standard model is included in the warped throat of a KKLT model, D-terms are naturally introduced. The idea is to include the standard model on a D3 (or anti-D3) brane localized at a fixed point at the tip of the throat. The blowing-up modes associated with the singularities combine with the axions needed for anomaly cancellation to give new complex moduli. These contribute to the 4D potential as a D-term. The resulting model has the many desired features: inflation, reheating, and the standard model. Numerical work is included in \cite{Burgess:2004kv} illustrating several possible trajectories for the inflaton. From this, it appears that it is much easier to get less than 60 e-folds of inflation in string theory, leading the authors to suggest some modifications of the usual inflation scenario (such as two periods of inflation). Other work relating D-term inflation to brane inflation includes \cite{Kallosh:2003ux} and \cite{Halyo:2003wd}.

The final class of terms, interbrane interactions, received the most attention prior to KKLT. The inter-brane potential was the focus of most earlier brane inflation work, when it was simply assumed that some mechanism fixed the moduli. It was found that in general brane/anti-brane scenarios have a potential that is too steep to give slow roll. However, there are at least three known solutions to this problem, all of which can be used in the KKLMMT context. The first has already been discussed: inflation between branes in a warped throat works because the warp factor can flatten the potential (see Eq.(\ref{DDbar})). The second is used in D3/D7 inflation, which takes advantage of the D-term scenario discussed in the previous section. The third possibility is to use branes at angles \cite{Jones:2002cv}, \cite{Buchan:2003gx}. Branes intersecting at multiple angles may provide a way to more naturally include realistic particle content (chiral fermions) \cite{Gomez-Reino:2002fs}. Then the inter-brane potential depends on the angle roughly as $\theta^3$ in the constant $A$ in Eq.(\ref{DDbar}). The potential may then be quite flat for small angles. Since T-duality relates the D3/D7 model to inflation between two sets of D5-branes at angles, this possibility may also be incorporated in the current discussion. Along with the D3/D7 model, this shares the advantage of not needing the warping, which simplifies calculations. Other work on $\DD$ inflation includes \cite{Buchel:2004qg}, \cite{Buchel:2003qj} and \cite{Blanco-Pillado:2004ns}.

Typically, we require the brane interaction term to give slow roll and then focus on the other effects that may ruin inflation. However, a recent proposal \cite{Iizuka:2004ct} takes advantage of Coulombic interactions to balance other contributions. The picture there is a compactification with symmetric throats containing anti-D3 branes. For a D3 placed near the symmetry point, the inter-brane interactions cancel the contribution from the superpotential and the system then satisfies slow-roll. 

\subsection{Complete models and observation}
The models discussed in \S2 are not quite complete since they do not address the issue of stabilization of Kahler moduli. In \cite{Denef:2004dm}, a series of examples is found that completely realizes the procedure outlined in KKLT. That is, the dilaton and complex structure are fixed with fluxes, meta-stable dS minima are available, and non-perturbative effects stabilize all the Kahler moduli. Mathematical requirements limit the models that can do all this, but among those that work are Fano threefolds. These have the additional advantage that they have been well studied and classified \cite{Greene:1990ud}. The main result for these models is that the procedure for fixing all moduli works in many cases but not generically. The models are constructed such that $\alpha^{\prime}$ corrections can be ignored. There is a claim that mathematically any usable model must have more than one Kahler moduli, which is already a complication from the examples cited above, although \cite{Gorlich:2004qm} argues that this restriction is not necessary.

After checking that a particular model gives enough inflation and a reasonable spectrum of density perturbations, one can look for signatures that may distinguish string theory models. Details and consequences of D3/D7 inflation are considered in \cite{Dasgupta:2004dw}. A detailed analysis of the results from a KKLT scenario, including a possible explanation of dark energy, is found in \cite{Garousi:2004uf}. Cosmic strings as a consequence of brane inflation are developed in \cite{Sarangi:2002yt} and \cite{Jones:2003da}. Work on cosmic strings as D- and F-strings includes \cite{Copeland:2003bj}, \cite{Dvali:2003zj}, \cite{Jackson:2004zg}, and \cite{Polchinski:2004hb}. The possibility that the cosmic strings formed may be semilocal is discussed in \cite{Dasgupta:2004dw} and \cite{Urrestilla:2004eh}.

\section{Conclusions}
We have seen that in practice the question of slow-roll in stringy inflation models can become mired in detailed calculations, especially in realistic compactifications. In spite of this, we would like to know if slow-roll can be expected as a general feature. The simplest approach seems to be to look for shift symmetries: flat directions that preserve slow-roll regardless of the results of the more difficult calculations. Our main conclusion is that the evidence so far (from $D3-\bar{D3}$ and $D3-D7$ models), indicates that shift symmetry, at least to first order, is generic and so we expect slow-roll to be possible.

From our sample calculations, we also see that T-duality is a useful check on the form of the K\"{a}hler potential. It may be interesting to use T-duality in a more complete model to find, for example, the dual description of the D3/D7 model as D5 branes at angles. We can also imagine extending this idea to see what can be found from checking mirror-symmetric descriptions. At the very least, we will learn something about moduli stabilization since we would exchange moduli that are currently stabilized by very different (perturbative vs. non-perturbative) methods.

Finally, we note that there are recent ideas for achieving inflation that do not rely on the types of calculations discussed here. For example, \cite{Silverstein:2003hf} and \cite{Chen:2004gc} make use of the higher order kinetic terms in the DBI action and a ``speed limit'' on the moduli space to find enough inflation without the need for a flat potential. This model has distinctive observational signatures which are discussed in \cite{Alishahiha:2004eh}. In general, it predicts more non-Gaussianities in the CMB spectrum and more power in tensor modes than the usual scenarios. Ideally, observations will allow us to compare competing models.

\vspace{0.3cm}

{\bf Acknowledgements}
I thank John Hsu, Renata Kallosh, Louis Leblond, Liam McAllister, Alex Saltman, Sash Sarangi and especially Hassan Firouzjahi and Henry Tye for valuable discussions.

\end{document}